\documentclass[twocolumn,aps,showpacs,prl,superscriptaddress]{revtex4-1}
\usepackage[latin9]{inputenc}
\usepackage{amsmath}
\usepackage{amssymb}
\usepackage{graphicx}
\usepackage{esint}
\usepackage[T1]{fontenc}
\PassOptionsToPackage{normalem}{ulem}
\usepackage{ulem}
\usepackage{pdfpages}

\usepackage{times}

\makeatletter
\@ifundefined{textcolor}{}
{%
 \definecolor{BLACK}{gray}{0}
 \definecolor{WHITE}{gray}{1}
 \definecolor{RED}{rgb}{1,0,0}
 \definecolor{GREEN}{rgb}{0,1,0}
 \definecolor{BLUE}{rgb}{0,0,1}
 \definecolor{CYAN}{cmyk}{1,0,0,0}
 \definecolor{MAGENTA}{cmyk}{0,1,0,0}
 \definecolor{YELLOW}{cmyk}{0,0,1,0}
}

\usepackage{color}%\usepackage{psfrag}%To include LaTeX subscripts on figures

\makeatother

\begin{document}

\title{Sudden Expansion of a One-Dimensional Bose Gas from Power-Law Traps}
\author{A.~S.~Campbell}
\affiliation{School of Physics and Astronomy, University of Birmingham, Edgbaston,
Birmingham, B15 2TT, United Kingdom}
\author{D.~M.~Gangardt}
\affiliation{School of Physics and Astronomy, University of Birmingham, Edgbaston,
Birmingham, B15 2TT, United Kingdom}
\author{K.~V.~Kheruntsyan}
\affiliation{The University of Queensland, School of Mathematics and Physics,
Brisbane, Queensland 4072, Australia}

\begin{abstract}
  We analyze free expansion of a trapped one-dimensional Bose gas after a
  sudden release from the confining trap potential. By using the stationary
  phase and local density approximations, we show that the long-time
  asymptotic density profile and the momentum distribution of the gas are
  determined by the initial distribution of Bethe rapidities (quasimomenta)
  and hence can be obtained from the solutions to the Lieb-Liniger
  %Bethe ansatz 
  equations in the thermodynamic limit. For expansion from a harmonic trap, and in the limits of
  very weak and very strong interactions, we recover 
  the self-similar scaling solutions known from the hydrodynamic approach.
  %the known scaling solutions of the hydrodynamic approach corresponding to self-similar expansion. 
  For all other power-law traps and arbitrary interaction
  strengths, the expansion is not self-similar and shows strong dependence 
  of the density profile evolution on the trap anharmonicity.
  %shows strong dependence, on the trap anharmonicity, of the shape variation of the density profile during evolution. 
  We also characterize dynamical fermionization of the expanding
  cloud in terms of 
  %its first- and second-order coherences
  correlation functions describing phase and density fluctuations.
\end{abstract}


\date{\today}
\maketitle

The vast majority of natural and laboratory-induced phenomena occur 
in interacting many-particle systems that are away from the equilibrium. Yet 
the nonequilibrium dynamics of such systems remains an unsolved problem 
both in quantum physics and some areas of classical physics such as fluid dynamics.
Examples include quantum and classical turbulence, dynamics across phase transitions, and 
plasma instabilities, to name a few.  
Ultracold atomic gases have recently
emerged as a particularly promising platform to gain new insights into aspects
of nonequilibrium dynamics of quantum many-body systems \cite{Bloch2008,Lamacraft-chapter} and,
more generally, into nonequilibrium statistical mechanics. In particular,
there has been a surge of research activity in the study of the dynamics 
after a sudden quench of the system's parameters, exploring the mechanisms of relaxation and
the role of integrability in approaches to equilibrium
\cite{Kinoshita2006,Hofferberth2007,Rigol2007,*Rigol2008,Gangardt2008,Fleischhauer2010,Caux2012,*Causx2013,*Kormos-Caux-Imambekov2013,Bolech2012,Fabian-exp,*Fabian-theory,Wright2014} (for further references, see \cite{Cazalilla2010,Polkovnikov2011}).

In this Letter, we study far-from-equilibrium behavior of a trapped quantum
gas after a sudden quench of the confining potential. More specifically, we
investigate free expansion of an interacting one-dimensional (1D) Bose gas
instantaneously released from the confining trap potential $V(x)$.  This is a
paradigmatic example of a ``quantum explosion'' problem in an experimentally
realizable system that can be described by an integrable microscopic
model---the Lieb-Liniger model \cite{Lieb-Liniger-I} of delta-interacting
bosons in one dimension. The exact integrability of the model offers an
opportunity to investigate the expansion dynamics using theoretical methods
that would have otherwise been inapplicable. At the same time, 
integrability implies that the underlying system lacks any mechanism of
thermalization, which in turn poses a question of applicability of the
standard hydrodynamic approach that was previously used to describe the
dynamics of this system
\cite{Castin1996,Kagan1996,*Kagan1997,Menotti2002,Pedri2003,Fang2014}.
Combining these aspects together gives us a unique opportunity to:
(i) solve the quantum explosion problem in a nontrivial manner, and
(ii) benchmark the predictions of the hydrodynamic approach
against those obtained here.

In this work, we  treat the simplest case
of expansion from the zero-temperature ground state of the trapped gas.  
As we show below, the asymptotic
density and momentum distributions of the gas after a sufficiently long
expansion time (once the expansion becomes ballistic) 
can be obtained from the initial distribution of quasimomenta of the trapped
(nonuniform) gas using the stationary phase approximation. This promotes the
initial quasimomentum distribution from an auxiliary quantity---which has so
far only been used to derive thermodynamic quantities---to the status of
an observable physical property. The initial quasimomentum distribution itself
is  calculated by combining the exact solutions of the uniform Lieb-Liniger model
and the local density approximation.

To start, we consider Hamiltonian  evolution in free space of the
 many-body wave function, which---immediately prior to the removal of the
 trap potential at $t\!=\!0$---describes the
ground state of $N$ trapped particles  
and is expanded in terms of the eigenfunctions
of the uniform Lieb-Liniger model,
\begin{gather}
\Psi(x_{1},\ldots,x_{N};t)=\frac{1}{(2\pi)^{N/2}}\int dk_{1}\dots dk_{N}\nonumber \\
\times b(k_{1}, \dots,k_{N})\, e^{i\Theta(k_{1},\,\dots,\, k_{N})}e^{i\sum_{j}(k_{j}x_{j}-\hbar k_{j}^{2}t/2m)}.\label{eq:psi_t_M}
\end{gather}
Here, $b(k_{1},\dots,k_{N})$ are the expansion coefficients of the initial
wave function \cite{Sup}, which depend on $N$ different quasimomenta $\{k_{j}\}$ (also
referred to as rapidities, and having units of wave numbers) and are normalized to $\int dk_{1}\ldots
dk_{N}\;|b(k_{1},\ldots,k_{N})|^{2}=1$. The phase $\Theta(k_{1},\ldots,k_{N})
= \sum_{j<l}\tan^{-1}[\hbar^2(k_{l}-k_{j})/mg]$ arises from two-body 
collisions by the
diffractionless delta-function interaction potential with the strength $g$ \cite{g1D}.

The expansion coefficients $b(k_1,\dots, k_N)$
determine the joint $N$-particle
probability distribution of quasimomenta $|b(k_1,\dots,k_N)|^{2}$. In the single-particle sector, i.e., after integration
over all quasimomenta but one, they give the quasimomentum distribution of the trapped gas,
\begin{eqnarray}
g(k)=N\int dk_{2}\dots dk_{N}\,|b(k,k_{2},\dots,k_{N})|^{2},
\label{eq:dens_quasi_M}
\end{eqnarray}
with the normalization $\int\! g(k)dk\!=\!N.$

To proceed, we note that the only \textit{time-dependent} term in the integrand of Eq.~(\ref{eq:psi_t_M}) is the phase of the last exponential. This exponential will, for sufficiently long times, develop fast oscillations as functions of $k_j$ compared to the remaining \textit{time-independent} terms. Therefore, in the long-time and large-distance limit, the asymptotic form of the wave function can be simplified significantly by using the stationary phase approximation \cite{Jukic2008}:
the main contribution to the integral in Eq.~(\ref{eq:psi_t_M}) comes from the stationary phase points \cite{Sup} satisfying
\begin{equation}
\frac{\hbar k_{j}}{m}
=\frac{x_{j}}{ t},\label{eq:stat_phase_M}
\end{equation}
and leading to the following asymptotic wave function:
\begin{gather}
\Psi_{\infty}(x_{1},\dots,x_{N};t)=\left(\frac{m}{\hbar t}\right)^{N/2}b\left(\frac{mx_{1}}{\hbar t},\ldots,\frac{mx_{N}}{\hbar t}\right)\nonumber \\
\times e^{i\Theta\left(\frac{mx_{1}}{\hbar t},\,\dots,\,\frac{mx_{N}}{\hbar t}\right)+i\frac{m}{2\hbar t}\sum_{j}x_{j}^{2}}e^{-i\pi N/4}.\label{eq:psi_asy_M}
\end{gather}

The corresponding asymptotic density distribution
$\rho_{\infty}(x,t)\!=\!N\int\! dx_{2}\dots
dx_{N}|\Psi_{\infty}(x,x_{2},\dots,x_{N};t)|^{2}$ can therefore be found,
using Eq.~(\ref{eq:dens_quasi_M}), as
\begin{eqnarray}
\rho_{\infty}(x,t) = \frac{m}{\hbar t}\; g\left(\frac{mx}{\hbar t}\right),
\label{eq:density_M}
\end{eqnarray}
with $\int\!  \rho_{\infty}(x,t)dx\!=\!N$. Thus, the density profile after long
expansion time is determined by the rescaled shape of the initial quasimomentum distribution
$g(k)$; finding this distribution constitutes, therefore, a key task of the present work.

The asymptotic wave function (\ref{eq:psi_asy_M}) also determines 
the asymptotic momentum distribution of the gas. Indeed, the Fourier
transform $\widetilde{\Psi}_{\infty}(k_{1},\dots,k_{N};t)$ of
Eq. (\ref{eq:psi_asy_M})
is again dominated by the stationary
phase points satisfying Eq.~(\ref{eq:stat_phase_M}), regarded now as
conditions on the positions $x_j$.  The result is
\begin{equation}
\widetilde{\Psi}_{\infty}(k_{1},\ldots,k_{N};t)\!=\!
(-i)^{N}b(k_{1},\ldots,k_{N})e^{i\Theta(k_{1},\dots,k_{N})}. 
\label{eq:psy_asy_k_res_M}
\end{equation}
Integrating $|\widetilde{\Psi}_{\infty}(k_{1},\ldots,k_{N};t)|^{2}$ over all
momenta but one and using Eq.~(\ref{eq:dens_quasi_M}), one obtains the asymptotic momentum distribution,
\begin{eqnarray}
n_{\infty}(k,t)=g(k),
\label{eq:dens_mom_M}
\end{eqnarray}
implying that the initial quasimomenta of the trapped gas are mapped to real 
momenta of the expanded cloud \cite{Jukic2008,Bolech2012}.
Then, the result of Eq.~(\ref{eq:density_M}) for the density profile 
can simply be viewed as a consequence of the ballistic 
position-momentum correlations, Eq.~(\ref{eq:stat_phase_M}), established 
in the long-time asymptotic regime, after the interaction energy has converted into
the kinetic energy of expanding particles.

The characteristic expansion time $t_\mathrm{e}$ ensuring the
applicability of the stationary phase approximation can be estimated by
requiring that the fastest particles, moving with the velocity 
$v_\mathrm{max}\sim c_0$, where $c_0$ is  the speed of sound, 
overtake all slower particles. This is equivalent to 
$c_0t_\mathrm{e}$ becoming larger than the characteristic size $R$ of
the initial cloud, and hence $t_\mathrm{e}\! \sim\! R/c_0$. 
For expansion from a harmonic trap with frequency $\omega_0$ this yields $t_\mathrm{e}\sim 1/\omega_0$ 
in the Thomas-Fermi approximation, in both the weakly and strongly interacting regimes (see below).

Finding the coefficients $b(k_{1},k_2\dots,k_{N})$ and 
hence the distribution $g(k)$, Eq.~(\ref{eq:dens_quasi_M}), is equivalent to
solving for the ground state of the (nonintegrable) trapped gas and  
 thus constitutes a formidable task for systems with
  large $N$  \cite{3particles}.
However, as we argue here, for large $N$, the quasimomentum
distribution $g(k)$ can be found approximately within the local density
approximation (LDA) \cite{nk_LDA}.

The LDA is invoked by assuming that the initial trapped cloud can be
divided into subsystems of length $\Delta L$ small compared its overall
size $R$, so that the density
$\rho(x)\!\equiv\!\rho(x,t\!=\!0)$ within each subsystem centered at $x$ is
approximately constant. At the same time $\Delta L$ has to be sufficiently
large compared to the microscopic correlation length $\xi_0$ so that the locally uniform subsystem can be treated via the 
solution of the Lieb-Liniger integral equation \cite{Lieb-Liniger-I} in the thermodynamic limit. Detailed conditions for the applicability of the LDA to trapped 1D Bose gases have been discussed in Ref.~\cite{Kheruntsyan2005}, in addition to being verified experimentally \cite{Amerongen2008,*Armijo2011,*Jacqmin2011}. Generally speaking, the LDA is expected to be very good in the bulk of the atomic cloud for sufficiently large $R$ (much larger than the length scale associated with the trapping potential), breaking down only in the small vicinity ($\sim \xi_0$) of the cloud edge.

The solution to the Lieb-Liniger integral equation for each region gives the local quasimomentum distribution
$f(k,x;t=0)$ \cite{Lieb-Liniger-I} corresponding to density $\rho(x)$, obtained for the local
value of the chemical potential $\mu(x)\!=\!\mu_{0}-V(x)$ \cite{Kheruntsyan2005}, where $\mu_{0}$ is the
global chemical potential. Integrating $f(k,x;t=0)$ over $x$ will give the quasimomentum distribution of the trapped gas in the local 
density approximation,
\begin{equation}
g(k)  = \int f(k,x;0)\, dx,
\label{eq:dens_quasi_2_M}
\end{equation}
whereas $\rho(x) \!=\!  \int\! f(k,x;0) dk$, 
with the normalization condition $\int\! f(k,x;0)dkdx\!=\!N$.

\begin{figure}[tbp]
\includegraphics[width=8.0cm]{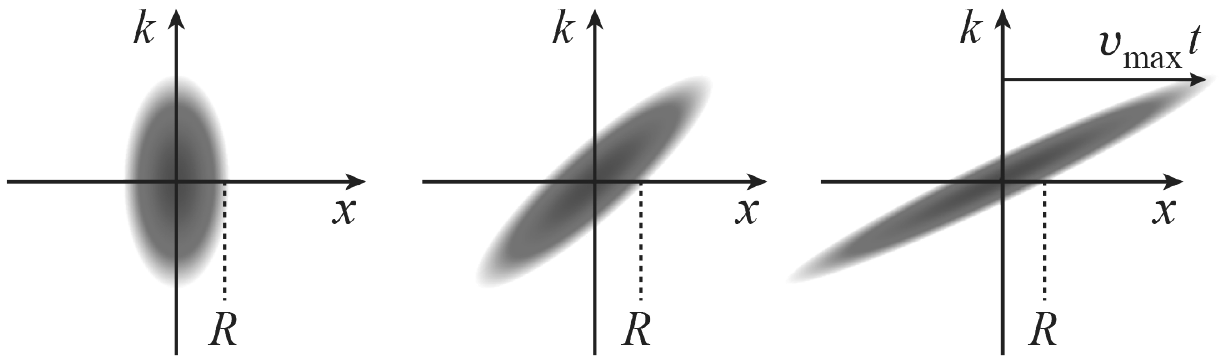}
\caption{Illustration of the evolving local quasimomentum distribution
  $f(k,x;t)\!=\!f(k,x\!-\!\hbar k t/m ;0)$, where $v_{\max}\!=\!\hbar k_{\max}/m$.
}
\label{fig:fkxt} 
\end{figure}

It is insightful to see how  the  asymptotic density evolution
  Eq.~(\ref{eq:density_M})  can  be obtained directly from the local
  quasimomentum  distribution $f(k,x;t)$ if the latter is provided with a
  semiclassical time dependence of the form
$f(k,x;t)\!=\!f(k, x\!-\!\hbar kt/m;0)$, for $t\!\gg \!t_{\mathrm{e}}$. This
choice makes use of
the ballistic expansion relationship $x\!=\!\hbar kt/m$ of
Eq.~(\ref{eq:stat_phase_M}) and is illustrated in  Fig.~\ref{fig:fkxt}.
It is clear that such an evolution 
leaves the (quasi)momentum distribution intact, 
\begin{eqnarray}
  \label{eq:gkt}
 g(k,t)=\! \int \! f(k,x-\hbar k t/m; 0)\; dx  =  g(k), 
\end{eqnarray}
 whereas the density distribution evolves according to  
 \begin{eqnarray}
   \label{eq:rhokt}
 \rho(x,t)\!=\!\!\int\! \!f(k,x;t)\;dk=\!\!\int\! \!f\!\left(k,x-\hbar k t/m;0\right)dk.  
\end{eqnarray}

Introducing a new variable $y\!\equiv \!x-\hbar kt/m$, Eq. (\ref{eq:rhokt}) can be
rewritten as $\rho(x,t)\!=\!\frac{m}{\hbar t}\int \!f\left(\frac{mx}{\hbar
    t}-\frac{my}{\hbar t},y;0\right)dy$, where we can further neglect
$my/\hbar t$ in the first argument as the main contribution to the integral
comes from the values of $y$ that are of the order of the initial
size of the cloud $R$ and values of $x\!\propto \!t$ that are much larger than
$R$ in the long-time limit. Using 
Eq.~(\ref{eq:dens_quasi_2_M}), one then obtains the same result as in
Eq.~(\ref{eq:density_M}),
$\rho_{\infty}(x,t)\!=\! \frac{m}{\hbar t}\int dy f\left(\frac{mx}{\hbar
    t},y;0\right)\!=\!\frac{m}{\hbar t} g\left(\frac{mx}{\hbar t}\right)$,
    as anticipated.

We now apply our approach to expansion from power-law traps,
$V(x)\!=\!\frac{1}{2}\alpha_{\nu}|x|^{\nu}$, where $\nu\!\ge\! 2$ and $\alpha_{\nu}$
is the confinement strength (in the important case of 
a harmonic trap, $\nu\!=\!2$ and $\alpha_2\!=\!m\omega_0^2$).  
The problem can be treated analytically in two
limiting cases: a weakly interacting gas ($\gamma_{0}\ll1$) and a strongly
interacting gas in the Tonks-Girardeau (TG) regime ($\gamma_{0}\!\rightarrow \!\infty$),
where $\gamma_{0}\!=\! mg/(\hbar^{2}\rho_{0})$ is the dimensionless interaction
strength in the trap center, with $\rho_{0}$ being the peak density of the
initial trapped sample. The intermediate regime can be addressed 
by finding numerically the local quasimomentum distribution $f(k,x;0)$ via
the solution of the Lieb-Liniger integral equation \cite{Lieb-Liniger-I} 
and then using Eqs.~(\ref{eq:dens_quasi_2_M}) and (\ref{eq:density_M}).
The results
of such a numerical treatment for a harmonic trap are shown in
Fig.~\ref{fig:examples}~(a), whereas the analytic results (see below) for
$\nu\!\neq \!2$ are illustrated in Figs.~\ref{fig:examples}(b) and 2(c).

\begin{figure}[tbp]
\includegraphics[width=7.3cm]{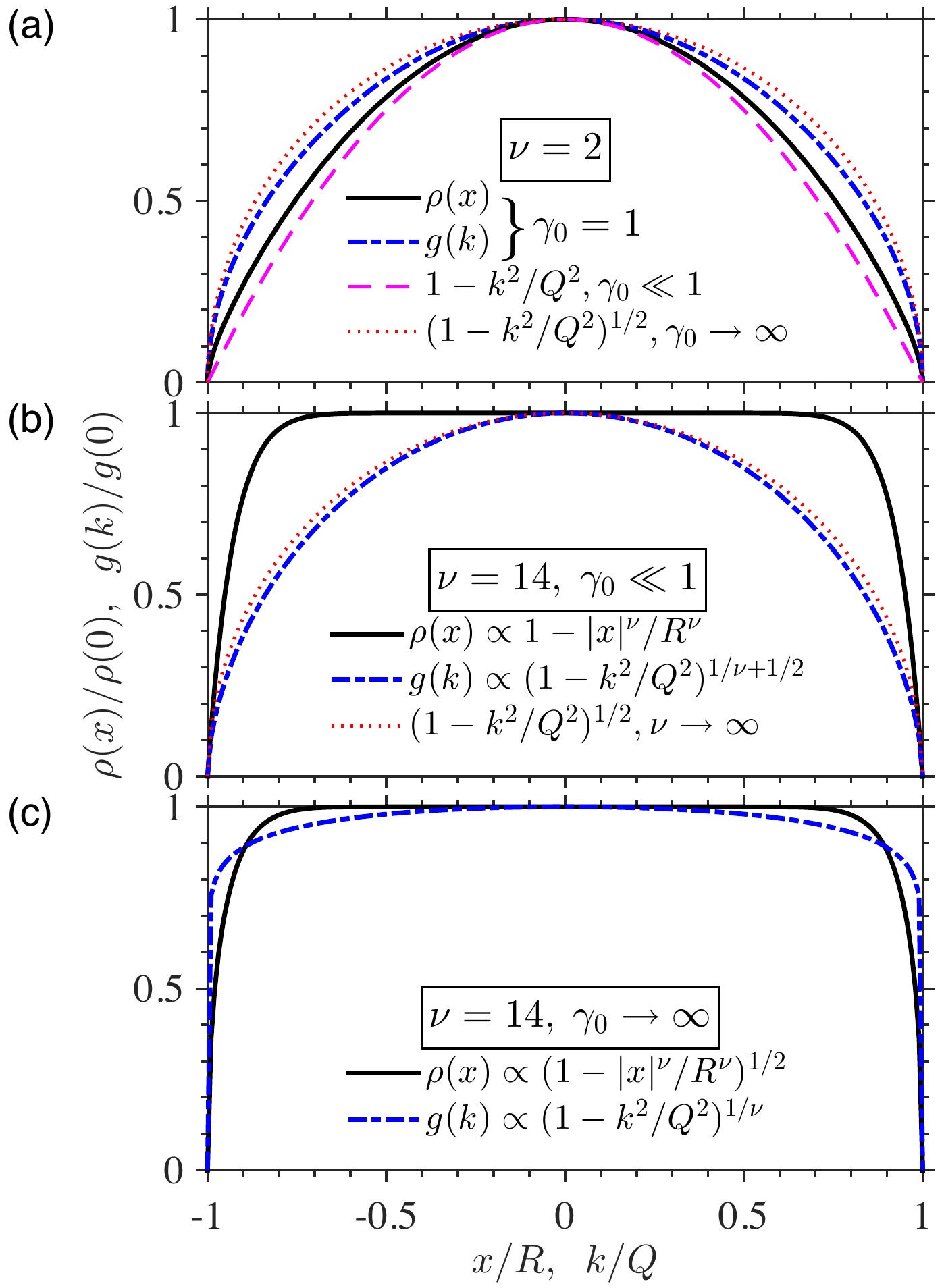}
\caption{(Color online) Examples of the initial density profile $\rho(x)$
  (black solid lines) and the quasimomentum distribution $g(k)$ (blue
  dash-dotted lines), which determines the shape of the asymptotic density
  profile $\rho_{\infty}(x,t)$ and the momentum distribution
  $n_{\infty}(k,t)$, Eqs.~(\ref{eq:density_M}) and (\ref{eq:dens_mom_M}),
  with $Q$ being the maximum (quasi)momentum. (a) Expansion from
  a harmonic trap, with the solid and dash-dotted lines corresponding to
  the numerical results for $\gamma_0\!=\!1$. The analytic results
  (coinciding with the hydrodynamic self-similar solutions) in the
  weakly ($\gamma_0\!\ll \!1$) and strongly ($\gamma_0\!\to
  \!\infty$) interacting  regimes are shown, respectively, by the dashed-magenta and 
  dotted-red lines. The numerical results in these regimes, obtained for
  $\gamma_0\!=\!2.5 \!\times\! 10^{-4}$ and $\gamma_0\!=\!200$, are
  indistinguishable from the respective analytic curves and are omitted from
  the graphs for clarity. (b) Main curves are for a highly
  anharmonic trap with $\nu\!=\!14$ \cite{boxtrap} and $\gamma_0\!\ll \!1$;
  the semicircle (dashed red line) corresponds to the limiting behavior of
  $g(k)$ for a box potential ($\nu\!\to\!\infty$). (c)  Strongly interacting regime, for the same $\nu\!=\!14$.  }
\label{fig:examples} 
\end{figure}

In the weakly interacting regime ($\gamma_0\!\ll\! 1$), 
the local semiclassical distribution of a trapped gas with the density $\rho(x)$ 
is given by \cite{Lieb-Liniger-I}
\begin{equation}
f(k,x;0)=
\frac{1}{2\pi}\sqrt{\frac{4\hbar^{2}\rho(x)}{mg}-\frac{\hbar^{4}k^{2}}{m^{2}g^{2}}},\; \text{for}\; |k|\!<\!K(x),
\label{eq:f_kx_GP_M}
\end{equation}
and $f(k,x;0)\!=\!0$ otherwise, with $K(x)\!=\!\sqrt{4mg\rho(x)/\hbar^{2}}$.

The equation of state for a uniform gas in this regime is 
$\rho=\mu/g$, and therefore the density profile of the trapped sample
in the Thomas-Fermi limit is given by
\begin{equation}
\rho(x)=\mu(x)/g=\rho_{0}\left(1-|x|^{\nu}/R^{\nu}\right),\; \text{for}\; |x|\!<\!R,
\label{eq:rho_GP_M}
\end{equation}
and $\rho(x)\!=\!0$ otherwise. Here, $R\!=\!
(2\mu_{0}/\alpha_{\nu})^{1/\nu}$
is the Thomas-Fermi
radius, and
$\rho_{0}\!=\!\mu_{0}/g\!=\![(1+\nu)^{\nu} \alpha_{\nu}N^{\nu}/(2^{\nu+1}\nu^{\nu}g)]^{1/(1+\nu)}$
is the peak density found from
the normalization condition $N\!\!=\!\!\int\!\rho(x)dx\!=\!2R\rho_{0}\nu/(1+\nu)$.

Integrating the distribution function $f(k,x;0)$, Eq. (\ref{eq:f_kx_GP_M}),
over position gives
\begin{align}
  g(k)
  =\frac{RI_{\nu}}{\pi}\left(\frac{4\hbar^{2}\rho_{0}}{mg}\right)^{1/2}\left(1-\frac{\hbar^{2}k^{2}}{4mg\rho_{0}}\right)^{1/2+1/\nu},\label{eq:g_GP_gen_M}
\end{align}
for $|k|\!<\!\sqrt{4mg\rho_{0}/\hbar^{2}}$, and $g(k)=0$ otherwise. Here,
 $I_{\nu}\! \equiv \!\int_{0}^{1}dy\,(1-y^{\nu})^{1/2}\!=\!\frac{\sqrt{\pi}\,\Gamma(1/\nu)}{(2+\nu)\,\Gamma(1/2+1/\nu)}$,
with $\Gamma(z)$ being the gamma function. 
The asymptotic density distribution, from Eq.~(\ref{eq:density_M}),
is then determined by 
\begin{equation}
\rho_{\infty}(x,t)=\frac{4I_{\nu}}{\pi}\frac{\rho_{0}}{\lambda
  (t)}\left(1-\frac{x^{2}}{\lambda  (t)^{2}R^{2}}\right)^{1/2+1/\nu},\label{eq:rho_GP_lambda_M}
\end{equation}
where we have introduced a dimensionless parameter 
$\lambda (t)=t/t_\mathrm{e} = 2c_0 t / R$, with $c_0 = \sqrt{\mu_0/m}$ being the sound velocity in the trap center. 
By comparing this result with the initial density distribution, Eq.~(\ref{eq:rho_GP_M}), 
it is now easy to see that for $\nu\!=\!2$, for which
$I_2\!=\!\pi/4$ and $t_\mathrm{e} =
1/(\sqrt{2}\omega_0)$, we
immediately reproduce the scaling solution of
Refs.~\cite{Castin1996,Kagan1996}, in which $\lambda (t)=\sqrt{2}\omega_0 t$
takes the meaning of the single scaling parameter. In the hydrodynamic
approach, $\lambda(t)$ is obtained from the scaling equation
$\ddot{\lambda}=\omega_0^2/\lambda ^2$ \cite{Sup}.  We can also immediately
conclude that such a self-similar scaling solution is not supported by any
other power-law trap potential. In particular, the case of 
$\nu=14$ illustrated in Fig.~\ref{fig:examples}~(b)
shows a dramatic difference between the density profiles of 
the initial and expanded clouds (\textit{cf.} \cite{Hadzibabic2013}).

In the TG regime ($\gamma_0\!\to \!\infty$), 
the local semiclassical distribution $f(k,x;0)$ is given by 
\cite{Lieb-Liniger-I}
\begin{equation}
f(k,x;0)=1/2\pi, \;\text{for}\; |k|<\pi \rho(x),
\label{eq:fkx-Tonks_M}
\end{equation}
and $f(k,x;0)\!=\!0$ otherwise, where $\pi\rho(x)$ is the maximum quasimomentum coinciding with the 
Fermi momentum of an ideal uniform Fermi gas of density $\rho(x)$.

The density profile $\rho(x)$ is found from the equation of state
of a uniform system $\rho=\sqrt{2m\mu/(\hbar\pi)^{2}}$,
yielding
\begin{equation}
\rho(x)=\sqrt{2m\mu(x)/(\hbar\pi)^{2}}=\rho_{0}\sqrt{1-|x|^{\nu}/R^{\nu}},
\label{eq:rho_TG_M}
\end{equation}
for $|x|\!<\!R,$ and $\rho(x)\!=\!0$ otherwise. Here, $R=(2\mu_{0}/\alpha_{\nu})^{1/\nu}$ 
is the Thomas-Fermi radius and $\rho_{0}=[2m\mu_{0}/(\hbar^{2}\pi^{2})]^{1/2}=[m\alpha_{\nu}N^{\nu}/(2^{\nu}I_{\nu}^{\nu}\hbar^{2}\pi^{2})]^{1/(2+\nu)}$ is the peak density, with 
$I_{\nu}$ being the same numerical coefficient as in
Eq. (\ref{eq:g_GP_gen_M}).

Integrating $f(k,x;0)$, Eq. (\ref{eq:fkx-Tonks_M}), over position gives 
\begin{equation}
g(k)=\frac{R}{\pi}\left(1-\frac{k^{2}}{\pi^{2}\rho_{0}^{2}}\right)^{1/\nu},\label{eq:g_TG_gen}
\end{equation}
for $|k|<\pi\rho_{0}$, and $g(k)=0$ otherwise. 
The asymptotic density distribution is therefore given by
\begin{equation}
\rho_{\infty}(x,t)=\frac{\rho_{0}}{\lambda (t)}\left(1-\frac{x^{2}}{\lambda (t)^{2}R^{2}}\right)^{1/\nu},
\label{eq:rho_TG_lambda_M}
\end{equation}
where $\lambda (t)=t/t_\mathrm{e} =  c_0 t /R$ and $c_0 = \sqrt{2\mu_0/m}$ is the sound (Fermi) velocity  
in the trap center. By comparing $\rho_{\infty}(x,t)$ with the initial density distribution, Eq.
(\ref{eq:rho_TG_M}), we immediately see that, for finite $\nu$ (see also \cite{Comment3}), a self-similar scaling solution
is again supported only by a quadratic potential, in which case 
$\lambda (t)=\omega_0 t$. In the hydrodynamic approach,
this asymptotic behavior is obtained from the scaling equation $\ddot{\lambda}
=\omega_0^2/\lambda ^3$ \cite{Sup} (see also \cite{Kolomeisky2001}), which also follows from the exact
treatment of Ref. \cite{Minguzzi:2005}.

Considering now the coherence properties of an expanding 1D Bose gas, we note
that the only length scale entering into the asymptotic momentum distribution
$n_{\infty}(k,t)$ [through Eqs.~(\ref{eq:dens_mom_M}) and either
(\ref{eq:g_GP_gen_M}) or (\ref{eq:g_TG_gen})] and hence into the respective
one-body density matrix $G^{(1)}(x,x';t)$ is the \textit{microscopic}
correlation length $\xi_0\!=\!\hbar/\sqrt{m\mu_0}$ \cite{Kheruntsyan2005,Sykes2008}, corresponding to  
the healing length $\xi_0\!=\!\hbar/\sqrt{mg\rho_0}$ for the weakly interacting
gas and the mean interparticle separation $\xi_0\!=\!1/\rho_0$ for the TG gas.
In all cases, $\xi_0$ is much smaller than the size of the sample $R$, 
which implies complete loss of phase coherence (if there was any
initially) typical of fermions
and can be viewed as a manifestation of ``dynamical
fermionization'' discussed in Refs.~\cite{Rigol2005,Minguzzi:2005,*Buljan2008}. Such a loss of
phase (or first-order) coherence with expansion, which we note does not follow
from the hydrodynamic approach, is indeed the case for a weakly interacting
gas, for which the initial (zero-temperature equilibrium) coherence length
$l^{(0)}_{\phi}\!\sim\! \xi_0e^{2\pi/\sqrt{\gamma_0}}$ \cite{Petrov2000} is
exponentially large and can typically be much larger than $R$.

Another manifestation of dynamical fermionization during expansion can be
seen in the asymptotic behavior of the same-point two-body correlation
function
$g^{(2)}(x,x;t)$: it acquires \cite{Sup} a scaling $\propto \!1/t^2$, 
implying suppressed correlation in the long time limit.
Such a suppression indicates dynamical 
approach to the fermionized TG regime, where 
$g^{(2)}(x,x)\!=\!0$ due to an effective Pauli exclusion \cite{Gangardt2003a,Kheruntsyan2003}. 
Moreover, our dynamical result can be 
written as $g^{(2)}(x,x;t)\!\propto \!1/\gamma(x,t)^2$ using the inverse 
scaling of the
instantaneous interaction constant $\gamma(x,t)\!\equiv
\!mg/\hbar^2\rho_{\infty}(x,t)$ with density 
$\rho_{\infty}(x,t)\!\propto\! 1/t$. Such a scaling of the $g^{(2)}$-function with
$\gamma$ is indeed typical of an equilibrium TG gas
\cite{Gangardt2003a,Kheruntsyan2003}. It must be noted though that this result
is by no means an indication of equilibration during expansion as the time
scale that establishes the $1/\gamma(x,t)^2$ scaling is still given by
$t_\mathrm{e}$ independently of the initial interaction strength; it is true
for even an initially weakly interacting gas with $\gamma(x,0)\!\ll \!1$ and
can emerge long before the instantaneous value of the interaction strength
itself becomes ``fermionic'', $\gamma(x,t)\!\gg \!1$, due to its own scaling of
$\gamma(x,t)\!\propto \!t$.

In summary, we have analyzed the far-from-equilibrium dynamics of the
Lieb-Liniger gas in a quantum explosion scenario of a sudden expansion from
the confining trap potential. Considering a general class of power-law traps, we have found  
the asymptotic density profiles and the momentum
distributions of the expanding clouds using the stationary phase and local density approximations.
The expansion is generally not
self-similar, except for the strongly and 
weakly interacting gases released from a quadratic trap for which our results are in agreement with the 
known hydrodynamic scaling solutions. In all cases, the expanding clouds
lose their phase coherence and display fermionic density fluctuations (not
accounted for by the hydrodynamic theory) on a time scale $t_\mathrm{e}\! \sim\!R/c_0$ 
by which the expansion becomes ballistic.
The zero-temperature results presented here are qualitatively valid 
for $k_BT\!\ll \!\mu_0$, however,
our approach can be easily generalized to a nonzero-temperature initial
state using the Yang-Yang approach \cite{Yang1969}, in which case the 
role of the initial local quasimomentum distribution $f(k,x;t=0)$ will be taken by its 
temperature-dependent counterpart to be found as in Refs. \cite{Kheruntsyan2003,Kheruntsyan2005} 
from the solutions to the Yang-Yang integral equations.

The authors acknowledge stimulating discussions with M. Pustilnik and
I. Bouchoule, and support by the ARC Discovery Project
DP140101763. 

%\bibliography{expansion}

\begin{thebibliography}{50}
\bibitem [{3pa()}]{3particles}%
  \BibitemOpen
  \href@noop {} {}\bibinfo {note} {For systems with few particles, the problem
  can be solved numerically and has been considered previously for $N=3$
  \cite{Jukic2008}, for the case of expansion from a quadratic
  trap.}\BibitemShut {Stop}%
\end{thebibliography}

\onecolumngrid\newpage



\section*{Supplementary material}

# Supplemental Material:
# Sudden expansion of a one-dimensional Bose gas from power-law traps

A. S. Campbell,[1] D. M. Gangardt,[1] and K. V. Kheruntsyan[2]

[1]*School of Physics and Astronomy, University of Birmingham, Edgbaston, Birmingham, B15 2TT, UK*
[2]*The University of Queensland, School of Mathematics and Physics, Brisbane, Queensland 4072, Australia*

## I. THE MANY-BODY WAVE FUNCTION

The Hamiltonian time evolution of the $N$-particle wave function $\Psi(x_1, x_2, \ldots, x_N; t)$ describing the dynamics of the Lieb-Liniger gas in free space, after its sudden release at time $t=0$ from the confining trap potential, can be written down using an expansion in eigenfunctions $\Psi_{\{k_j\}}(x_1, x_2, \ldots, x_N)$ of the uniform Lieb-Liniger model,

$$\Psi(x_1, x_2, \ldots, x_N; t) = \frac{1}{\sqrt{N!}} \int dk_1 \ldots dk_N$$
$$\times b(k_1, \ldots, k_N) \Psi_{\{k_j\}}(x_1, x_2, \ldots, x_N) e^{-iE_{\{k_j\}}t}. \quad (S1)$$

Here, the eigenfunctions $\Psi_{\{k_j\}}(x_1, \ldots, x_N)$ are characterized by $N$ different quasimomenta $\{k_j\} = k_1, k_2, \ldots, k_N$ with eigenenergies $E_{\{k_j\}} = \sum_{j=1}^{N} \hbar^2 k_j^2/2m$, whereas the expansion coefficients $b(k_1, \ldots, k_N)$ depend on the initial trapping potential and are normalized to $\int dk_1 \ldots dk_N |b(k_1, \ldots, k_N)|^2 = 1$. The eigenfunctions $\Psi_{\{k_j\}}(x_1, \ldots, x_N)$ in the fundamental region $x_1 < x_2 < \ldots < x_N$ can be written down in explicit form as a sum of $N!$ partial waves [1],

$$\Psi_{\{k_j\}}(x_1, \ldots, x_N) = \frac{1}{\sqrt{(2\pi)^N N!}}$$
$$\times \sum_P (-1)^P e^{i\Theta(k_{P_1}, \ldots, k_{P_N})} e^{i\sum_j k_{P_j} x_j}, \quad (S2)$$

enumerated by permutations $P$. Each of the $N!$ terms in the sum is multiplied by the sign of the permutation defined to be $(-1)^P = \pm 1$ for an even/odd permutation, *i.e.*, a permutation which can be obtained from the identity permutation $\{k_j\} = k_1, k_2, \ldots, k_N$ by an even/odd number of transpositions of adjacent elements. The phase

$$\Theta(k_1, \ldots, k_N) = \sum_{j<l} \frac{1}{2i} \log \frac{1 + i\frac{\hbar^2(k_l - k_j)}{mg}}{1 - i\frac{\hbar^2(k_l - k_j)}{mg}}$$
$$= \sum_{j<l} \tan^{-1}\left[\frac{\hbar^2(k_l - k_j)}{mg}\right] \quad (S3)$$

arises from two-body scattering by the diffractionless delta-function potential of strength $g$ (see main text).

The structure of the eigenstates in Eq. (S2) has the following physical interpretation: in the fundamental region $x_1 < x_2 < \ldots < x_N$, choosing the decreasing arrangement of quasi-momenta $k_1 > k_2 > \ldots > k_N$ in Eq. (S2) implies that the identity permutation term, in which $P_j = j$ ($j = 1, 2, \ldots N$), corresponds to $N$ particles entering the collision, with the faster particles being behind the slower particles in consecutive order from left to right. The term corresponding to the permutation obtained from the identity by swapping the indices $i$ and $j$ describes the scattering of particles with quasi-momenta $k_i$ and $k_j$ and therefore acquires a scattering phase $\theta(k_i - k_j) = 2\tan^{-1}[\hbar^2(k_i - k_j)/mg]$. Proceeding in a similar fashion one generates all other terms in the expansion (S2); in particular, the maximally crossed permutation term with $P_j = N - j + 1$ corresponds to $N$ particles leaving the collision, with the faster particles having overtaken the slower ones.

Rearranging the quasimomenta $\{k_Q\} = k_{Q_1}, k_{Q_2}, \ldots, k_{Q_N}$ does not change the physical situation; however, the wave-function (S2) acquires the factor $(-1)^Q$. Therefore, in order to avoid double counting of the physical states in the $N$-particle bosonic wavepacket (S1), while at the same time ensuring that it remains symmetric, one requires that the coefficients $b(k_1, \ldots, k_N)$ are antisymmetric functions of quasimomenta.

Substituting Eq. (S2) into Eq. (S1) and using the antisymmetry property of the expansion coefficients $b(k_1, \ldots, k_N)$ gives a particularly simple expression for the initial state of the trapped Lieb-Liniger gas,

$$\Psi(x_1, x_2, \ldots, x_N; 0) = \frac{1}{(2\pi)^{N/2}} \int dk_1 \ldots dk_N$$
$$\times b(k_1, \ldots, k_N) e^{i\Theta(k_1, \ldots, k_N)} e^{i\sum_j k_j x_j}, \quad (S4)$$

valid in the fundamental region $x_1 < x_2 < \ldots < x_N$. In any other region the symmetry of the bosonic functions under permutation of particles should be used. Equation (S4) describes the initial $N$-body wavepacket; its time evolution after a sudden removal of the confining potential is governed by the free Lieb-Liniger Hamiltonian and is given by

$$\Psi(x_1, x_2, \ldots, x_N; t) = \frac{1}{(2\pi)^{N/2}} \int dk_1 \ldots dk_N$$
$$\times b(k_1, \ldots, k_N) e^{i\Theta(k_1, \ldots, k_N)} e^{i\sum_j (k_j x_j - \hbar k_j^2 t/2m)}, \quad (S5)$$

which is the same as Eq. (1) of the main text.

To evaluate the integral in Eq. (S5) using the stationary phase approximation, we note that the main contribution to the integral comes from the stationary phase points $k_j^* = mx_j/\hbar t$ corresponding to $d\phi(k_j)/dk_j = 0$, where $\phi(k_j) \equiv k x_j - \hbar k_j^2 t/2m$. The integral is then calculated by approximating the slowly varying function $b(k_1, \ldots, k_N) e^{i\Theta(k_1, \ldots, k_2)}$ by its value at the stationary points and pulling it outside the integral; the remaining integral over the exponentials of phases $\phi(k_j)$ are evaluated by using the Taylor expansion near

the points $k_j^*$ and keeping only the first two nonzero terms, $\phi(k_j) \approx \phi(k_j^*) + \frac{1}{2}\phi''(k_j^*)(k_j - k_j^*)^2$. As a result, we obtain

$$\Psi_\infty(x_1, x_2, \ldots, x_N; t) = \left(\frac{m}{\hbar t}\right)^{N/2} b\left(\frac{mx_1}{\hbar t}, \ldots, \frac{mx_N}{\hbar t}\right)$$
$$e^{i\Theta\left(\frac{mx_1}{\hbar t}, \ldots, \frac{mx_N}{\hbar t}\right) + i\frac{m}{2\hbar t}\sum_j x_j^2} e^{-i\pi N/4}, \quad (S6)$$

which coincides with Eq. (4) of the main text.

## II. TWO-BODY CORRELATION FUNCTION

The calculation of the two-body local (same point) correlation function,

$$G^{(2)}(x, x; t) = \frac{N(N-1)}{2}$$
$$\times \int dx_3 \ldots dx_N \left|\Psi_\infty(x, x, x_3, \ldots, x_N; t)\right|^2, \quad (S10)$$

requires the knowledge of the wavefunction at two coinciding points. The asymptotic wavefunction Eq. (S6) must, in this case, be modified to take into account the zero of the prefactor

$$G(k_1, k_2, \ldots, k_N) \equiv b(k_1, k_2, \ldots k_N) e^{i\Theta(k_1, k_2, \ldots k_N)} \quad (S11)$$

at $k_1 = k_2 = mx/\hbar t$. Expanding this prefactor around the stationary points $k_j^* = mx_j/\hbar t$ to the second order, one obtains the required asymptotic wavefunction,

$$\Psi_\infty(x, x, x_3, \ldots, x_N; t) = \frac{m}{i\hbar t}\left(\frac{m}{\hbar t}\right)^{N/2} e^{-i\pi N/4}$$
$$\times e^{i\frac{m}{2\hbar t}\sum_j x_j^2} \sum_{j=1}^N \left.\frac{\partial^2 G}{\partial k_j^2}\right|_{k_j = k_j^*}. \quad (S12)$$

This form of the wavefunction manifests the effective 'fermionization' via the additional factor $1/t$ as compared to Eq. (S6). Substituting Eq. (S12) into Eq. (S10) and using a variable change $y_j = mx_j/\hbar t$ for $j = 3, \ldots, N$ results in the following scaling of the pair correlation function with time: $G^{(2)}(x, x; t) \propto 1/t^4$. Normalizing this with the product of densities $\rho_\infty(x, t)$, yields

$$g^{(2)}(x, t) = \frac{G^{(2)}(x, x; t)}{\rho_\infty(x, t)^2} \propto \frac{1}{t^2}, \quad (S13)$$

which is the result discussed in the main text, indicating the 'dynamical fermionization' of expanding bosons.

## III. COMPARISON WITH THE HYDRODYNAMIC APPROACH

As shown in the main text, the free expansion of the Lieb-Liniger gas from a quadratic ($\nu = 2$) trap leads to self-similar scaling solutions in the limits of very weak and very strong interactions. Such scaling solutions can be obtained from the hydrodynamic approach [2–5], which we would like to illustrate here for instruction.

In the hydrodynamic approach, the evolution of the density and velocity fields, $\rho(x, t)$ and $v(x, t)$, under the variations of the trap frequency $\omega(t)$, is governed by the following coupled equations (see, e.g., [4, 5]),

$$\partial_t \rho + \partial_x(\rho v) = 0,$$
$$\partial_t v + v\partial_x v = -\partial_x\left(\frac{1}{2}\omega(t)^2 x^2\right) - \frac{1}{m\rho}\partial_x P, \quad (S14)$$

where $P$ is the pressure.

In the weakly interacting regime ($\gamma_0 \ll 1$), the pressure is given by $P(x, t) = g\rho(x, t)^2/2$, which follows from the equation of state $\mu = g\rho$ and the Gibbs-Duhem thermodynamic relation $\rho = (\partial P/\partial \mu)_T$. Using this and substituting the scaling ansatz for $\rho(x, t)$ from Eq. (15) of the main text for $\nu = 2$ into the above hydrodynamic equations, yields, for $\omega(t) = 0$ (i.e., for complete opening of the confining trap), the following equation for the scaling parameter $\lambda(t)$:

$$\ddot{\lambda} = \omega_0^2/\lambda^2. \quad (S15)$$

The initial conditions are $\lambda(0) = 1$ and $\dot{\lambda}(0) = 0$. Multiplying Eq. (S15) by $\dot{\lambda}$ one obtains the integral of motion $\dot{\lambda}^2 + 2\omega_0^2/\lambda = \text{const} = 2\omega_0^2$, which gives the asymptotic solution $\lambda(t) = \sqrt{2}\omega_0 t$ for $t \to \infty$, i.e., the same result as the one discussed in the main text.

In the Tonks-Girardeau regime ($\gamma_0 \to \infty$), the pressure is given by $P(x, t) = \hbar^2\pi^2\rho(x, t)^3/3m$, from the equation of state $\mu = \hbar^2\pi^2\rho^2/2m$. In this case, substituting the scaling ansatz (19) of the main text (for $\nu = 2$) into the hydrodynamic equations with $\omega(t) = 0$ gives the following equation for the scaling parameter:

$$\ddot{\lambda} = \omega_0^2/\lambda^3. \quad (S16)$$

The integral of motion is now $\dot{\lambda}^2 + \omega_0^2/\lambda^2 = \text{const} = \omega_0^2$, which gives $\lambda(t) = \omega_0 t$ for $t \to \infty$, in agreement with what we obtained in the main text. This result can also be obtained [6] within the exact treatment of the Tonks-Girardeau gas using time-dependent Bose-Fermi mapping and the scaling transformation applicable to the time-dependent Schrödinger equation for the single-particle orbitals.



\end{document}